\documentclass[aps,prd,
preprintnumbers,nofootinbib,superscriptaddress]{revtex4}
\usepackage[dvips]{graphicx}
\usepackage{caption}
\usepackage{ragged2e}
\usepackage{subcaption}
\usepackage{xcolor}
\usepackage{bm,latexsym,amsmath,amssymb,amsfonts,color}
\usepackage{ragged2e}
\usepackage{array}
\usepackage{float}

\usepackage{ragged2e}
\begin{document}

\title{Robustness of exact $SU(N)$ Einstein-Skyrme solutions: \\ black holes, wormholes and cosmologies}

\author{Carla Henr\'iquez-Baez}
\email{carla.henriquez@umayor.cl}
\affiliation{Centro Multidisciplinario de F\'isica, Vicerrector\'ia de Investigaci\'on, Universidad Mayor, Camino La Pir\'amide 5750, Santiago, Chile.}
\affiliation{Grupo de Investigación en Física Teórica, GIFT, Universidad Católica de la Santísima Concepción, Alonso de Ribera 2850, Concepción, Chile.}

\author{Perla Medina}
\email{perla.medina
@ulagos.cl}
\affiliation{Departamento de Ciencias Exactas, Universidad de Los Lagos, Av. Fuchslocher 1305, Osorno,
Chile.}

\author{Pablo Rodr\'iguez}
\email{pablo.rodriguez@ulagos.cl}
\affiliation{Departamento de Ciencias Exactas, Universidad de Los Lagos, Av. Fuchslocher 1305, Osorno,
Chile.}

\author{Aldo Vera}
\email{aldo.vera@umayor.cl}
\affiliation{N\'ucleo de Matem\'atica, F\'isica y Estad\'istica, Universidad Mayor, Avenida Manuel Montt 367, Santiago, Chile}
\affiliation{Centro Multidisciplinario de F\'isica, Vicerrector\'ia de Investigaci\'on, Universidad Mayor, Camino La Pir\'amide 5750, Santiago, Chile.}

\begin{abstract}
The $SU(2)$ Skyrme model admits exact solutions when coupled to General Relativity in four spacetime dimensions. Here, we demonstrate that relevant gravitational configurations, such as black holes, wormholes, and cosmologies, are robust under the extension to $SU(N)$ with an arbitrary number of flavors. We show that key physical features, such as the thermodynamic properties, the baryon number of the configurations, and the geometry, are crucially determined by the number of flavors as well as the coupling constants of the theory. In particular: first, we prove that an increase in the matter content favors the local thermal stability of spherically symmetric black holes. Second, for cosmological solutions, we demonstrate analytically that as the hadronic matter content increases, the scale factor grows more slowly due to enhanced gravitational attraction, while the expansion rate (measured by the Hubble parameter) increases as a consequence of the higher energy density. Third, we find that the throat of traversable wormhole solutions can be enlarged by increasing the number of flavors, since the energy density required to keep the throat open does not grow indefinitely but rather saturates at a finite value.
\end{abstract}

\maketitle

\newpage

\tableofcontents

%%%%%%%%%%%%%%%%%%%%%%%%%%%%%%%%%%%%%%%
\section{Introduction}   \label{Sec-I}
%%%%%%%%%%%%%%%%%%%%%%%%%%%%%%%%%%%%%%%

The Skyrme model is a primary low-energy effective nonlinear field theory that provides a robust framework for describing hadronic matter \cite{Skyrme1, Skyrme2}. Based on a purely bosonic action, it captures the dynamics of both the bosonic and fermionic sectors; within this framework, mesons emerge as small fluctuations, while baryons are identified as topological solitons arising from the nonlinear interactions \cite{Witten:1979kh,Witten:1983tw,Witten:1983tw2}. These solitons, known as skyrmions, successfully characterize the static properties of nucleons when the internal symmetry group is $SU(2)$, representing the usual two-flavor case \cite{ANW} (see Refs. \cite{MantonBook, WeinbergBook, BalaBook} for comprehensive reviews). 

The study of Skyrme fields in curved geometries has been investigated since the pioneering works in Refs.  \cite{Manton:1986pz,Gibbons:2003cp}. Coupling the Skyrme model to gravity opens the possibility of exploring various astrophysically relevant configurations, including hairy black holes, compact stars, and cosmic evolutionary scenarios (see, for instance, Refs. \cite{Luckock:1986tr,Droz:1991cx,Heusler:1992av,Glendenning:1988qy,Bizon1,Bizon2,Kunz,Kunz2,Piette:2007wd,Nelmes,BD,BHRT}). In this context, the case of General Relativity in $D=4$ spacetime dimensions is of particular significance. By unifying these two well-established frameworks, it is expected that the theory provides a suitable setting for comparing analytical predictions with high-precision observational data from missions such as NICER, gravitational wave observatories like LIGO/Virgo, or pulsar timing arrays \cite{LIGOScientific:2017vwq,Riley:2019yda,Hobbs:2009yy}.

In addition, while it is well established that the Skyrme model provides a low-energy effective description of Quantum Chromodynamics (QCD) in the large $N_c$ limit \cite{tHooft:1973alw,Witten:1983tw2}, concrete setups are frequently restricted to two flavors. However, QCD involves additional flavors that play a critical role in several physical processes, such as hyperon formation and the physics of strange matter \cite{Callan:1985hy,Kaplan:1986yq}. Although incorporating these extra degrees of freedom is physically well-motivated, it typically induces a substantial increase in the complexity of the field equations \cite{Bala1,Bala2,Ioa,Brihaye:2003qs,Sato:2008zza}. To circumvent this technical challenge, the maximal embedding of $SU(2)$ into $SU(N)$, implemented within the framework of generalized Euler angles \cite{Bertini:2005rc,Cacciatori:2012qi,CacciatoriScotti} (see also Refs. \cite{Tilma1,Tilma2}), provides a highly efficient structure for exploring configurations with $N>2$, where $N$ denotes the flavor number, without the computational burden usually associated with additional fields.

Strikingly, despite the high degree of nonlinearity inherent in the Einstein-Skyrme field equations, the coupled system admits exact analytical solutions, a feature that is as mathematically remarkable as it is physically profound. In the past few years, an array of exact configurations describing black holes, extended objects, and gravitating solitons has been successfully constructed \cite{CanforaMaeda,Eloy1,Canfora:2017ivv,toroidal1,toroidal2,Eloy2,Canfora:2025roy,Canfora:2026col} (see also Refs.~\cite{gravtube1,gravtube2,Canfora:2026wfi,Canfora:2026dym,Canfora:2026ola}). These solutions, encompassing spherical, flat, toroidal, and string-like geometries, demonstrate that the flavor number $N$ non-trivially alters the thermodynamic properties and geometry of the system \cite{toroidal3, toroidal4}. Furthermore, extending self-gravitating Skyrmions to an arbitrary number of flavors via the maximal embedding Ansatz reveals that the flavor number explicitly determines the topological charge of the configuration \cite{Vera:2025qqz}. Moreover, the phenomenology associated with these novel exact solutions is consistent with the behavior previously uncovered in numerical studies of self-gravitating Skyrmions and hairy black holes \cite{Luckock:1986tr,Droz:1991cx,Glendenning:1988qy,Heusler:1992av,Bizon1} (see also Ref.~\cite{Bizon2}).

These developments highlight the ongoing interest in understanding the nonlinear features of hadronic matter coupled to gravity, thereby providing a strong motivation to explore multi-flavor extensions for diverse gravitational backgrounds, from dynamical collapse scenarios to static and stationary backgrounds. 

In this paper, we demonstrate that key self-gravitating configurations—namely black holes, wormholes, and cosmological backgrounds—remain robust under the inclusion of an arbitrary number of flavors, $N$. By systematically exploring the coupled Einstein-Skyrme system beyond the two-flavor restriction, we show how essential physical features, such as thermodynamic properties, stability criteria, and the expansion dynamics of the universe, are directly influenced by both the flavor number and the coupling constants of the theory. Ultimately, our findings establish that the addition of flavors provides a powerful mechanism to modulate the structural and dynamical properties of these configurations.

This article is organized as follows. In Sec.~\ref{Sec-II}, we briefly review the $SU(N)$ Einstein-Skyrme model. In Sec.~\ref{Sec-III}, we present exact spherical black hole solutions supported by the Skyrme field, compute their thermodynamic quantities, and discuss their stability. In Sec.~\ref{Sec-IV}, we present cosmological solutions and analyze their expansion rates. In Sec.~\ref{Sec-V}, we construct a wormhole solution whose traversable throat is critically determined by the number of flavors. Finally, Sec.~\ref{Sec-VI} presents our concluding remarks.

%%%%%%%%%%%%%%%%%%%%%%%%%%%%%%%%%%%%%%%
\section{Preliminaries}   \label{Sec-II}
%%%%%%%%%%%%%%%%%%%%%%%%%%%%%%%%%%%%%%%

\subsection{The Einstein-Skyrme model}

In this subsection, we briefly review the Skyrme theory coupled to Einstein gravity in $D=4$ spacetime dimensions.

The $SU(N)$ Einstein-Skyrme theory is described by the action 
\begin{gather}
    I[g,U]=\int d^4 x \sqrt{-g}\left(\frac{R-2\Lambda}{2\kappa} +\frac{K}{4}\text{Tr}\left[L^{\mu}L_{\mu}+\frac{\lambda}{8}G_{\mu \nu}G^{\mu \nu} \right] \right) \ , \label{act1} \\
    L_{\mu}=L_{\mu}^{j}t_{j} =U^{-1}\nabla_{\mu}U \ , \qquad G_{\mu \nu}=\left[L_{\mu},L_{\nu} \right] \ . \notag
\end{gather}
where $R$ is the Ricci scalar, $g$ is the metric determinant, and $t_j$ are the generators of the $SU(N)$ group. Regarding the coupling constants, $\Lambda$ is the cosmological constant, $\kappa=8\pi G$ is the gravitational constant, and $K$ and $\lambda$ are positive coupling constants fixed experimentally. The limit $\lambda \rightarrow 0$ corresponds to the nonlinear sigma model (NLSM).

The Skyrme equations, obtained by varying the action with respect to the $U$ field, together with Einstein's equations, are
\begin{subequations}
\begin{gather}
    \nabla^{\mu}L_{\mu}+\frac{\lambda}{4}\nabla^{\mu}\left[L^{\nu},G_{\mu \nu} \right]=0 \ , \label{S-eq} \\
    R_{\mu \nu}-\frac{1}{2}g_{\mu \nu}R+\Lambda g_{\mu \nu}=\kappa T_{\mu \nu} \ , \label{E-eq}
\end{gather}
\end{subequations}
where the energy-momentum tensor of the Skyrme matter field is given by 
\begin{equation}
    T_{\mu \nu} = -\frac{K}{2} \text{Tr}\left[L_{\mu}L_{\nu}-\frac{1}{2}g_{\mu \nu}L^{\alpha}L_{\alpha}+\frac{\lambda }{4}\left(g^{\alpha \beta}G_{\mu \alpha}G_{\nu \beta}-\frac{1}{4}g_{\mu \nu}G_{\alpha \beta}G^{\alpha \beta} \right) \right] \ .
\end{equation}
The topological charge in the Skyrme model, which corresponds to the baryon number, is defined by 
\begin{equation} \label{B}
B=\frac{1}{24\pi^{2}}\int_{\Sigma} \rho_{\text{B}}\, \mathrm{d}V\ , \qquad \rho_{\text{B}}=\epsilon^{ijk}\text{Tr}\biggl[\left( U^{-1}\partial_{i}U\right) \left( U^{-1}\partial_{j}U\right) \left( U^{-1}\partial_{k}U\right) \biggr] \ ,
\end{equation}
where the integration is performed over a spacelike hypersurface $\Sigma$. In Eq. \eqref{B}, the indices $\{i, j, k\}$ run over the spatial directions of spacetime.

%%%%%%%%%%%%%%%%%%%%%%%%%%%%%%%%%%%%

\subsection{Maximal embedding and arbitrary flavors}

To construct exact self-gravitating configurations with an arbitrary number of flavors, we adopt the maximal embedding of $SU(2)$ into $SU(N)$ within the framework of generalized Euler angles \cite{Bertini:2005rc,Cacciatori:2012qi,CacciatoriScotti}. This approach maps the three standard generators of the $SU(2)$ group onto an irreducible representation of spin $j=(N-1)/2$ composed of $N\times N$ matrices. This algebraic setup allows for the derivation of non-embedded exact solutions; that is, configurations that cannot be reduced to a trivial embedding of $SU(2)$, making the explicit role of the flavor parameter $N$ manifest (see, for instance, Refs. \cite{Pedro2,SU(N)1,SU(N)2}). A conceptually related construction was pioneered in early works using the $SO(3)$ subgroup of $SU(3)$ to investigate di-baryon states in the Skyrme model; see Refs. \cite{Bala1,Bala2}.

In the generalized Euler angles parametrization, the matter field $U(x)\in SU(N)$ associated with this maximal embedding is expressed as
\begin{equation}\label{matter-ansatz}
    U=e^{F_{1}(x^{\mu}) T_{3}}e^{F_{2}(x^{\mu}) T_{2}}e^{F_{3}(x^{\mu}) T_{3}}\,,
\end{equation}
where the degrees of freedom $F_i(x^\mu)$ are arbitrary functions of the spacetime coordinates. The matrices $T_i$ (with $i=1,2,3$) act as the generators of a three-dimensional subalgebra of $\mathfrak{su}(N)$, defined explicitly in terms of the matrix units $(E_{i,j})_{mn}=\delta_{im}\delta_{jn}$ by
\begin{subequations}
\begin{gather}
    T_1=-\frac{i}{2}\sum_{j=2}^{N}\sqrt{(j-1)(N-j+1)}(E_{j-1,j}+E_{j,j-1})\,,\label{T1}\\
    T_2=\frac{1}{2}\sum_{j=2}^{N}\sqrt{(j-1)(N-j+1)}(E_{j-1,j}-E_{j,j-1})\,,\label{T2}\\
    T_3=i\sum_{j=1}^{N}\left(\frac{N+1}{2}-j\right)E_{j,j}\,.\label{T3}
\end{gather}
\end{subequations}
By construction, these generators satisfy the standard $\mathfrak{su}(2)$ commutation relations
\begin{equation}
    [T_a,T_b]=\epsilon_{abc}T_c\,.
\end{equation}
The non-trivial nature of this embedding into $SU(N)$ is explicitly revealed by evaluating the Dynkin index or the trace of the product of these generators, which yields
\begin{equation}\label{Tr}
    \text{Tr}(T_b T_c)=-\frac{1}{2}a_N\delta_{bc}\,,\qquad a_N=\frac{N(N^2-1)}{6}\,.
\end{equation}
Here, the normalization factor $a_N$ corresponds to the tetrahedral numbers, highlighting a direct dependence on the flavor parameter $N$. Consequently, choosing different values for the number of flavors yields distinct configurations characterized by a specific spin. As we will demonstrate in the subsequent sections, this scaling is of fundamental importance since both the topological charge and the energy density of the self-gravitating system are directly determined by the trace of these generators, and thus by the flavor number $N$. Moreover, we will see that the results reported here are fully compatible with those using the ansatz in Refs. \cite{Bala1,Bala2,Eloy2} for the case of the $SU(3)$ group.

%%%%%%%%%%%%%%%%%%%%%%%%%%%%%%%%%%%%%%%
\section{The spherical black hole} \label{Sec-III}
%%%%%%%%%%%%%%%%%%%%%%%%%%%%%%%%%%%%%%%

In this section, we present an exact spherical black hole solution in the $SU(N)$ Einstein-Skyrme model, which extends both the two-flavor Canfora-Maeda solution reported in Ref. \cite{CanforaMaeda} and the spherical black hole solution in the context of the $SU(N)$ Einstein-NLSM presented in Ref. \cite{toroidal3}.

%%%%%%%%%%%%%%%%%%%%%%%%%%%%%%%%%%%%%%%
\subsection{The exact solution}

It is possible to construct a spherical black hole in the $SU(N)$ Einstein-Skyrme theory by considering the metric ansatz
\begin{equation}\label{sp-bh-1}
    ds^2 =-f(r^{\prime})dt^{\prime 2} + \frac{d r^{\prime 2}}{f(r^{\prime})}+r^{\prime 2}\left( d\theta^2 +\sin^2{\theta} d\phi^2 \right) \, ,
\end{equation}
together with the matter field $U$, where the functions $F_{i}$ in Eq. \eqref{matter-ansatz} are given by
\begin{equation}\label{f-sp-bh-1}
    F_{1}(x^\mu)= -\phi \, , \qquad F_{2}(x^{\mu})= 2\theta \, , \qquad F_{3}(x^{\mu})= \phi \, ,
\end{equation}
with $0 \leq \theta < \pi$, $0 \leq \phi < 2\pi$, and $r'$ being the radial coordinate. Note that the matter field defined in Eq. \eqref{f-sp-bh-1} possesses two proportional degrees of freedom; thus, according to Eq. \eqref{B}, the baryon number of this solution is zero. Consequently, the solution is supported solely by the mesons of the theory.\footnote{We will see below that this is not the case for the cosmological solution and the wormhole, where the ansätze for the matter field allow for baryonic degrees of freedom.}

With the previous ansatz, the Skyrme equations in Eq. \eqref{S-eq} are identically satisfied, while the Einstein equations in Eq. \eqref{E-eq} are solved analytically by
\begin{equation}\label{fsp-bh-1}
    f(r^{\prime})= 1-\frac{2 \bar{m}}{r'}-\frac{\Lambda}{3}r'^2 + K\kappa a_{N}\left( \frac{\lambda}{2(r^{\prime})^{2}} -1 \right) \, .
\end{equation}

This solution generalizes the Canfora-Maeda black hole with Skyrme matter \cite{CanforaMaeda} to an arbitrary number of flavors $N$, while extending the $SU(N)$ spherical black hole of Einstein-NLSM \cite{toroidal3} by including the Skyrme contribution to the matter action. The Canfora-Maeda solution describes an asymptotically (A)dS spherically symmetric black hole endowed with a solid angle deficit and a Skyrme contribution to the metric function that closely resembles the Maxwell $\sim 1/r^{2}$ term in the Reissner-Nordström geometry. In the asymptotically flat limit and devoid of the Skyrme matter term, the solution reduces to the global monopole black hole geometry previously discussed by Gibbons in Ref. \cite{Gibbons:1990um}. Furthermore, setting $\bar{m}=\Lambda=\lambda=0$, one recovers the Barriola-Vilenkin global monopole spacetime discussed in Ref. \cite{Barriola:1989hx}. In the present case, we can see from Eqs. \eqref{sp-bh-1}-\eqref{fsp-bh-1} that the solid angle deficit, $4\pi K\kappa a_N$, is entirely determined by the nonlinear matter field through the coefficient $a_{N}$.

In the analysis of the function $f(r^{\prime})$, the presence of the cosmological constant only changes the asymptotic behavior. For simplicity, we set $\Lambda=0$. In this case, the solution has two horizons
\begin{equation}
    r^{\prime}_{\pm} = \frac{1}{1-K\kappa a_{N}}\left( \bar{m} \pm \sqrt{\bar{m}^2-\left( 1-K\kappa a_{N} \right)\frac{K\kappa a_{N}\lambda}{2}} \right) \, .
\end{equation}

\begin{figure}[H]
    \centering
    \includegraphics[scale=1.1]{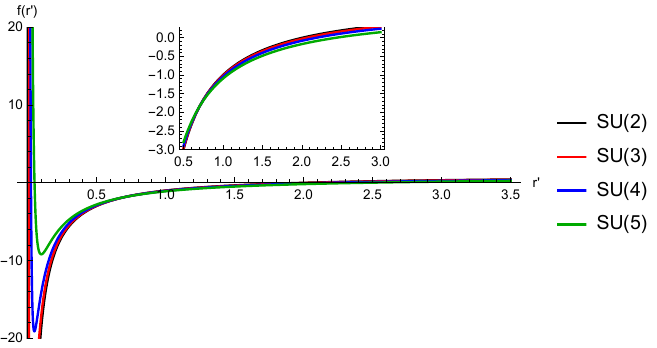}
    \caption{Behavior of the metric function $f(r^{\prime})$ for $\Lambda=0$, $\bar{m}=1$, $K=\kappa=0.1$, and $\lambda=1$. For clarity, an inset zoom is included highlighting the radius $r^{\prime}_{*} \sim 0.707$, where the hierarchy of the curves is reversed for this choice of parameters.}
    \label{f bh}
\end{figure}

In addition to the event horizons where $f(r^{\prime})=0$, Fig. \ref{f bh} shows the special radius $r^{\prime}_{*}=\sqrt{\lambda/2}$, where the hierarchy of the curves for different values of $N$ is reversed. At this radius, the contribution of the Skyrme term vanishes for any $N$, yielding $f(r^{\prime}_{*})=1-2\bar{m}/r^{\prime}_{*}$. The reversal in the curve hierarchy occurs because the Skyrme field contribution, $K\kappa a_{N}\left( \lambda/[2(r^{\prime})^{2}] - 1 \right)$, is positive for $r^{\prime}<r^{\prime}_{*}$, causing $f(r^{\prime})$ to increase with $N$, while for $r^{\prime}>r^{\prime}_{*}$, this term becomes negative and $f(r^{\prime})$ decreases with $N$. As mentioned above, introducing a cosmological constant only alters the asymptotic behavior, approaching $-\infty$ in the de Sitter case and $+\infty$ in the Anti-de Sitter case.

%%%%%%%%%%%%%%%%%%%%%%%%%%%%%%%%%%%%%%%
\subsection{Thermodynamics}

We are particularly interested in studying the local thermal stability of the spherical black hole and how it depends on the number of flavors $N$, an aspect analyzed below via the heat capacity. Accordingly, the AdS case provides a natural framework for this analysis, as black holes can be maintained in stable thermal equilibrium with a thermal bath. Hence, we restrict the following discussion to solutions with a negative cosmological constant, $\Lambda=-3/\ell^{2}$. To derive the thermodynamic properties of these black holes, we follow the regularized Euclidean action formalism developed in Ref.\cite{Flores-Alfonso:2019aae} for the $SU(2)$ Einstein-Skyrme theory. Within this framework, the metric in Eq. \eqref{sp-bh-1} is analytically continued to its Euclidean version by taking an imaginary time, $t\rightarrow i\tau$, and identifying its period $\beta$ with the inverse temperature of the black hole. To avoid conical singularities at the horizon, the Euclidean time period is fixed by $\beta=4\pi/f'(r_{+})$. Once the regularity of the Euclidean section has been imposed, the regularized on-shell Euclidean action is evaluated as 
\begin{equation}
I_{\text{E}}=I_{\text{gravity}}+I_{\text{matter}} \, , \label{IE}
\end{equation}
where $I_{\text{gravity}}$ comprises the bulk contribution, the Gibbons-Hawking boundary term, and the standard Anti-de Sitter counterterm required to render the gravitational action finite \cite{Emparan:1999pm}. On the other hand, the matter sector of the regularized Euclidean action contains the Skyrme bulk contribution and a suitable surface term.\footnote{In the present spherically symmetric case, only the surface counterterm associated with the $SU(N)$ NLSM sector is required to render the Euclidean action finite. The additional surface term associated with the $\lambda \neq 0$ Skyrme contribution, introduced in Ref. \cite{toroidal5}, is unnecessary here.} It is worth noting that the Skyrme matter field in Eq. \eqref{f-sp-bh-1} is not affected by the analytic continuation.

Before evaluating the Euclidean continuation of the solution in Eqs. \eqref{sp-bh-1}-\eqref{fsp-bh-1} within the regularized action, it is convenient to perform the coordinate transformation
\begin{equation}
    \tau^{\prime}= \frac{\tau}{\sqrt{1-K\kappa a_{N}}} \, , \qquad r^{\prime}= r\sqrt{1-K\kappa a_{N}} \, , 
\end{equation}
which leads to a two-sphere radius rescaled by a factor $\sqrt{1-K\kappa a_{N}}$, imposing the restriction $1-K \kappa a_N>0$ on the parameters. Thereby, the solution explicitly exhibits the solid angle deficit discussed above, and the metric function in Eq. \eqref{fsp-bh-1} can be written in the standard Reissner-Nordström-AdS form (under the identification $Q \rightarrow q_{N}$), namely
\begin{equation}
    f(r)=1-\frac{2m_{N}}{r}+\frac{r^{2}}{\ell^{2}}+\frac{q_{N}^2}{r^2} \, ,
\end{equation}
where $m_{N}=\bar{m}\left(1-K\kappa a_{N}\right)^{-3/2}$ is the Nucamendi-Sudarsky mass \cite{Nucamendi:1996ac}, and the effective hair parameter is defined by
\begin{equation} \label{qN}
    q_{N}^{2}= \frac{K\kappa a_{N}\lambda}{2\left(1-K\kappa a_{N}\right)^2} \, . 
\end{equation}

Therefore, the regularized Euclidean action in Eq. \eqref{IE} and the Hawking temperature, $T=1/\beta$, reduce to
\begin{equation} \label{bh-tem}
    I_{\text{E}} = \frac{4\pi\beta}{\kappa} \left( 1-K\kappa a_{N} \right) \left( m_{N} - \frac{r^{3}_{+}}{\ell^{2}} + \frac{q^{2}_{N}}{r_{+}} \right) \, , \qquad T = \frac{3r^{4}_{+}+\ell^{2}\left( r^{2}_{+} -q_{N}^{2} \right)}{4\pi \ell^{2}r^{3}_{+}} \, .
\end{equation}

According to the expression above for the regularized Euclidean action, it is possible to compute the entropy and mass of the black hole through the standard formulae, $S=(\beta \partial_{\beta}-1)I_{\text{E}}$, and $M=\partial_{\beta}I_{\text{E}}$, obtaining
\begin{equation}\label{S-M}
    S = \left( 1- K\kappa a_{N} \right) \frac{8\pi^{2}r_{+}^{2}}{\kappa} \, , \qquad M=\left( 1- K\kappa a_{N} \right)\frac{8\pi m_{N}}{\kappa} \, .
\end{equation}
At this point, it is important to note the following facts:
First, the entropy of the black holes satisfies the Bekenstein-Hawking  quarter of the area law, that is $S=A/4G$, with $A=4\pi r_{+}^{2}\left( 1- K\kappa a_{N} \right)$. Second, the entropy of the black hole remains strictly positive according to $1>K\kappa a_{N}$. Third, the black hole mass coincides with the ADM mass, or more precisely, the Abbott-Deser-Tekin mass, considering the Anti-de Sitter background. Finally, it is reassuring to verify that the system satisfies the first law of black hole thermodynamics, $\delta M= T\delta S$.\\
In order to study the local stability of the solution, we compute the heat capacity $C=-\beta^{2}\partial^{2}_{\beta}I_{\text{E}}$, obtaining the following 
\begin{equation} \label{HeatCapacity}
    C = 2S \frac{3r^{4}_{+}+\ell^{2}\left( r^{2}_{+} - q_{N}^{2} \right)}{3r^{4}_{+}+\ell^{2}\left( 3q_{N}^{2} - r^{2}_{+} \right)} \, .
\end{equation}

If the temperature and entropy are positive, local thermodynamic stability is achieved whenever the heat capacity is positive. Since the numerator of \eqref{HeatCapacity} is proportional to the Hawking temperature, positivity of the latter implies that the sign of the heat capacity is entirely determined by its denominator. The latter vanishes at
\begin{equation}
r_{*,\pm}^{\,2}=\frac{\ell^2}{6}\left(1\pm\sqrt{1-\frac{36q_N^2}{\ell^2}}\right).
\end{equation}

Since the effective hair parameter \eqref{qN} is always positive, for $q_N<\ell/6$, the two roots $r_{*,\pm}$ are real and positive. In this case, the heat capacity is positive for the small-black-hole branch above extremality ($T=0$), with $r_+<r_{*,-}$, and for the large-black-hole branch, with $r_+>r_{*,+}$, while the intermediate branch $r_{*,-}<r_+<r_{*,+}$ has negative heat capacity and is therefore locally unstable. As $q_N$ increases, the two limiting radii approach each other and coincide at $q_N=\ell/6$, for which $r_{*,-}=r_{*,+}=\ell/\sqrt{6}$. For $q_N>\ell/6$, the denominator of (19) is positive for all $r_+$, and hence every solution in the physical branch is locally thermodynamically stable. This behavior coincides with that of the Reissner-Nordstrom-AdS black hole. However, unlike the electric charge of the Reissner-Nordstrom-AdS solution, $q_N$ is not independent, but here is fixed by the Skyrme couplings and the flavor number through Eq. \eqref{qN}. Consequently, within $K\kappa a_N<1$, enlarging $N$ increases $q_N$, thereby reducing the locally unstable region and eventually removing it once $q_N>\ell/6$.

%%%%%%%%%%%%%%%%%%%%%%%%%%%%%%%%%%%%%%%
\section{Cosmological solutions} \label{Sec-IV}
%%%%%%%%%%%%%%%%%%%%%%%%%%%%%%%%%%%%%%%

In this section, we present exact cosmological solutions of the $SU(N)$ Einstein-Skyrme model describing Bianchi IX universes. Such scenarios have been previously studied in Refs.~\cite{Eloy1,Canfora:2017ivv} for the two-flavor and three-flavor cases independently (see also Refs. \cite{Pavluchenko:2016gyd,Eloy2,Canfora:2019sxn}), where exact solutions and their dynamical properties were analyzed. Here, we extend this construction to an arbitrary number of flavors $N$; we study how the flavor number affects the cosmological evolution through a detailed analysis of the behavior of the scale factor and the Hubble parameter.

Although the field equations allow for the description of universes with positive, negative, and vanishing cosmological constants, we pay close attention to the de Sitter case due to its phenomenological relevance and richer dynamical structure.

%%%%%%%%%%%%%%%%%%%%%%%%%%%%%%%%%%%%%%%%
\subsection{Isotropic Bianchi IX cosmology}

We consider an isotropic Bianchi IX spacetime, described by the metric
\begin{equation} \label{Bianchi}
    ds^2 =-dt^2 +\frac{\rho(t)^2}{4}\left[\left(d\gamma +\cos\theta d\phi\right)^2 +d\theta^2 +\sin^2 \theta d\phi^2\right] \ ,
\end{equation}
with a single scale factor, $\rho(t)$, together with the functions in the maximal embedding ansatz for the matter field in Eq. \eqref{matter-ansatz},
\begin{equation}
    F_{1}\left(x^{\mu} \right)= \gamma \, , \qquad F_{2}\left(x^{\mu} \right)= \theta \, , \qquad F_{3}\left(x^{\mu} \right)= \phi \, ,\label{cosmo-F's}
\end{equation}
where $0 \leq \theta < \pi$, $0 \leq \phi < 2\pi$, and $0 \leq \gamma < 4\pi$.

One of the key characteristics of these configurations is that they possess a non-zero topological charge, thus describing baryon states. Moreover, the topological charge depends directly on the number of flavors considered in the theory. Indeed, evaluating the topological charge density from Eq. \eqref{B} for the ansatz defined in Eq. \eqref{cosmo-F's} yields
\begin{equation} \label{charge}
    \rho_{\text{B}} = \frac{3}{2} a_N \sin\theta \quad \implies \quad  
    B=\frac{1}{24 \pi^2} \int \rho_{\text{B}} \, \mathrm{d}\gamma \, \mathrm{d}\theta \, \mathrm{d}\phi = a_N \ . 
\end{equation}
Thus, the topological charge is directly determined by the normalization factor $a_N$ defined in Eq. \eqref{Tr}. For the lowest allowed flavor numbers, $N = 2, 3, 4, \dots$, the resulting baryon numbers are $B = 1, 4, 10, \dots$, corresponding to the sequence of tetrahedral numbers. The case $N=2$ reproduces the solution reported in Ref. \cite{Eloy1}, whereas $N=3$ describes the solution in Ref. \cite{Eloy2}. Consequently, for large values of $N$, this framework enables the description of multi-baryon cosmological scenarios.

With the above metric and matter field configurations, the Skyrme field equations are identically satisfied, and the Einstein sector reduces to the following set of second-order differential equations for the scale factor:
\begin{subequations}
\begin{gather}
    \frac{\dot{\rho}^2}{\rho^{2}} -\frac{\Lambda}{3} -\frac{K\kappa a_{N}\lambda}{2}\frac{1}{\rho^{4}} + \left(1 -\frac{K\kappa a_{N}}{2}\right) \frac{1}{\rho^{2}} =0 \, ,\label{Energy-constraint} \\
    \ddot{\rho}-\frac{\Lambda}{3}\rho +\frac{K\kappa a_{N}\lambda}{2}\frac{1}{\rho^3 }= 0 \, . \label{E-P}
\end{gather} 
\end{subequations}
The first equation is a vanishing energy constraint, while the second corresponds to a special case of the well-known Ermakov-Pinney (EP) equation; see Refs. \cite{Ermakov,Pinney}. The exact solutions\footnote{The solutions in Eq.~\eqref{cosmological solution} are obtained from the standard Pinney construction, $\rho(t)= \sqrt{\alpha u^2 +\beta v^2+2\gamma uv}$, where $u(t)$ and $v(t)$ are two linearly independent solutions of the associated linear equation, and the constants satisfy $\alpha\beta-\gamma^2=-b/W^2$, with $W=u\dot{v}-v\dot{u}$ being the Wronskian. The integration constants are conveniently parametrized in terms of the initial scale factor $\rho_0=\rho(0)$.} of the EP equation for $\Lambda>0$, $\Lambda<0$, and $\Lambda=0$ are given respectively by
\begin{subequations}
\begin{gather}
    \rho^2_{\text{dS}}(t)= \rho_{0}^{2} \cosh^{2}(\bar{\omega} t) - \frac{b}{2\rho_{0}^{2}\bar{\omega}^{2}} \sinh^{2}(\bar{\omega} t) + \frac{\sqrt{b/2}}{\bar{\omega}}\sinh(2\bar{\omega} t) \, , \qquad \Lambda>0 \, , \label{rho deSitter0 1}\\
    \rho^2_{\text{AdS}}(t) = \rho_{0}^{2} \cos^{2}(\omega t) -\frac{b}{2\rho_{0}^{2}\omega^{2}} \sin^{2}(\omega t) + \frac{\sqrt{b/2}}{\omega}\sin(2\omega t) \, , \qquad \Lambda<0 \, , \label{rho AdS}\\
    \rho_{\text{flat}}^{\,2}(t) = \rho_0^2+\sqrt{2b}\,t -\frac{b}{2\rho_0^2}t^2 \, , \qquad \Lambda=0 \, , \label{rho flat}
\end{gather}
\label{cosmological solution}
\end{subequations}
where $\rho_{0}=\rho(0)$ and $b = K\kappa a_{N}\lambda/2$. Note that $\bar{\omega}=\sqrt{\Lambda/3}$ for $\Lambda>0$ and $\omega=\sqrt{-\Lambda/3}$ for $\Lambda<0$.

Evaluating Eq. \eqref{rho deSitter0 1} in the energy constraint \eqref{Energy-constraint} at $t=0$ yields a quartic equation for the initial scale factor $\rho_{0}$. Since $b>0$, the two branches for $\rho_{0}^{2}$ are real provided that $(1-c)^{2}-2b\bar{\omega}^{2}\geq 0$, where $c = K\kappa a_{N}/2$, and both are positive provided that $1-c>0$. Therefore, both conditions are simultaneously satisfied if
\begin{equation}
   1-c -\sqrt{2b\bar{\omega}^{2}}\geq 0 \, ,\label{condition de-Sitter 1}
\end{equation}
which provides a consistency condition on the parameters of the solution and, in particular, on the allowed number of flavors $N$ through the dependence of $b$ and $c$ on $a_N$. It is also important to characterize the dynamical evolution associated with these two branches, $\rho_{0,\pm}^{2}$. The energy constraint also determines the possible turning points of the cosmological evolution. Setting $\dot{\rho}_{\text{dS}}=0$ in Eq.~\eqref{Energy-constraint}, we obtain two possible turning points in the expansion, $\rho_{*,\pm}^{2}$. Therefore, together with the previous consistency condition and provided that
\begin{equation}
   (1-c)^2-4b\bar{\omega}^{2}\geq 0 \, ,\label{turn}
\end{equation}
the two turning points are real and positive. The energy constraint then separates the evolution into two allowed regions, $0<\rho_{\text{dS}}\leq\rho_{*,-}$ and $\rho_{\text{dS}}\geq\rho_{*,+}$, where $\dot{\rho}_{\text{dS}}^{2}\geq0$, separated by a forbidden gap $\rho_{*,-}<\rho_{\text{dS}}<\rho_{*,+}$, where $\dot{\rho}_{\text{dS}}^{2}<0$.

The two branches of the initial scale factor lie in different allowed regions. When both conditions in Eqs. \eqref{condition de-Sitter 1} and \eqref{turn} are simultaneously satisfied, the two branches and the turning points are ordered as $\rho_{0,-}<\rho_{*,-}<\rho_{*,+}<\rho_{0,+}$. Since $\dot{\rho}_{\text{dS}}(0)>0$ for both branches, the branch $\rho_{0,-}$ initially expands within the lower allowed region until reaching $\rho_{*,-}$, where the expansion stops and the solution recollapses. On the other hand, the branch $\rho_{0,+}$ starts above the larger turning point and continues to expand, approaching an asymptotically de Sitter regime.

It is worth noticing that the static self-gravitating $SU(N)$ Skyrmion previously discussed in Ref. \cite{Vera:2025qqz} arises as the limiting case in which the two turning points coincide. Indeed, setting $\rho(t)=\rho_s$ in Eqs. \eqref{Energy-constraint} and \eqref{E-P} yields
\begin{equation}
    b=\bar{\omega}^{2}\rho_s^{4} \, , \qquad \rho_s^{2}=\frac{1-c}{2\bar{\omega}^{2}} \, ,
\end{equation}
and combining both relations leads to $(1-c)^2-4b\bar{\omega}^{2}=0$, which is precisely the saturation condition in Eq. \eqref{turn}. Therefore, the static case can be identified with the degenerate limit where $\rho_{*,-}=\rho_{*,+}=\rho_s$.

The flavor number directly affects this dynamical structure through $b$ and $c$, which both increase with $a_N=N(N^2-1)/6$. Within the parameter region considered above, increasing $N$ brings the two turning points closer together and shifts the forbidden gap towards smaller values of the scale factor.

\begin{figure}[h!]
    \centering
    \begin{subfigure}[b]{0.48\textwidth}
        \centering
        \includegraphics[width=\textwidth]{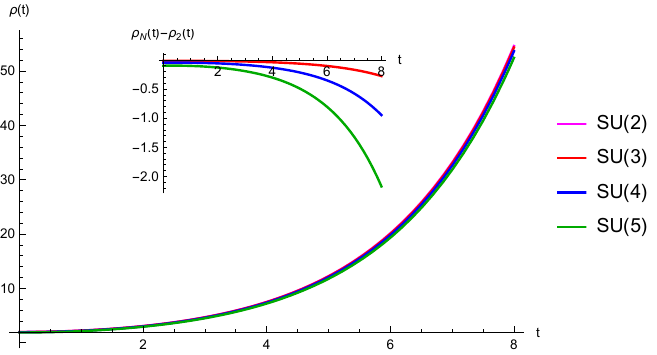}
        \caption{}
        \label{comparacion}
    \end{subfigure}
    \hfill
    \begin{subfigure}[b]{0.48\textwidth}
        \centering
        \includegraphics[width=\textwidth]{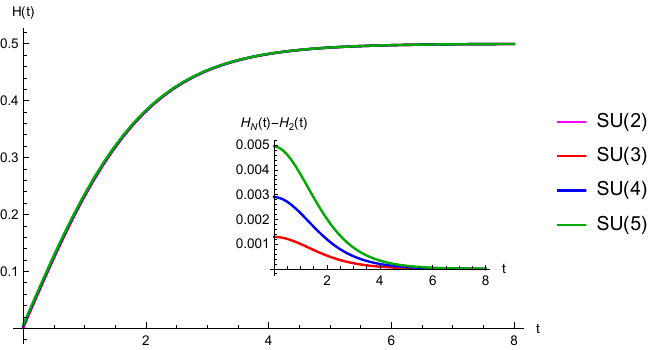}
        \caption{}
        \label{GraficoHubble}
    \end{subfigure}
    \caption{\justifying Behavior of the scale factor $\rho_{\text{dS}}(t)$ in (a) for $\Lambda=0.75$, $\lambda=0.01$, and $K\kappa=0.01$, considering the expanding branch $\rho_{0,+}$. In (b), the behavior of the Hubble function $H(t)\equiv\dot{\rho}_{\text{dS}}/\rho_{\text{dS}}$ is shown for the same parameters. An inset highlights the differences between $H_N(t)$ for $N=3,4,5$ and the $N=2$ case, which vanish asymptotically.}
    \label{dS}
\end{figure}

\begin{figure}[htbp]
    \centering
    \includegraphics[scale=0.9]{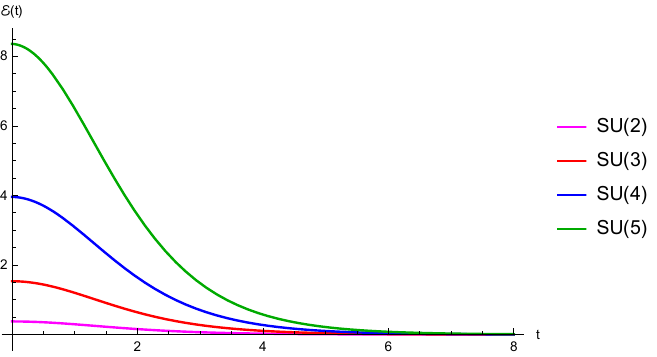}
    \caption{\justifying Energy density of the expanding branch $\rho_{0,+}$ for $\Lambda=0.75$, $\lambda=0.01$, and $K\kappa=0.01$. The energy density increases with the number of flavors $N$, while preserving the same qualitative decay in time.}
    \label{fig:Skyrme plus de-Sitter ED}
\end{figure}

Figs. \ref{comparacion} and \ref{GraficoHubble} show the behavior of the scale factor, $\rho_{\text{dS}}(t)$, and the Hubble function, $H=\dot{\rho}/\rho$. As expected for $\Lambda>0$, the branch $\rho_{0,+}$ exhibits an asymptotically de Sitter accelerated expansion for all the displayed values of $N$. As the number of flavors increases, the scale factor grows more slowly due to enhanced gravitational attraction, while at finite times, the expansion rate (measured by the Hubble parameter) increases as a consequence of the higher energy density (see Fig. \ref{fig:Skyrme plus de-Sitter ED}). This behavior can be understood from Eq. \eqref{Energy-constraint}, where $H^2$ contains the term $K\kappa a_{N}\lambda/(2\rho^{4})$, which grows linearly with $a_{N}$. Note that this phenomenon depends crucially on the Skyrme term and is absent for matter content supported solely by the NLSM model. Likewise, from the dynamical Eq. \eqref{E-P}, we see that the magnitude of the term proportional to $\rho^{-3}$ grows with $a_{N}$, acting as a gravitational deceleration in the expansion of the scale factor. Nevertheless, the effect of $N$ on the Hubble function is transient. Indeed, in the asymptotic regime, the expanding branch behaves as $\rho_{\text{dS}}(t)\sim C_N e^{\bar{\omega}t}$, where $C_N$ contains the dependence on the number of flavors. Therefore,
\begin{equation}
    \lim_{t\rightarrow\infty}H(t) = \bar{\omega} = \sqrt{\frac{\Lambda}{3}} \, ,
\end{equation}
independently of $N$. Hence, the number of flavors modifies the transient cosmological evolution, while the asymptotic expansion rate is entirely determined by the cosmological constant.

For $\Lambda<0$, the situation is less rich than in the de Sitter case. Evaluating the solution in Eq. \eqref{rho AdS} under the energy constraint at $t=0$, the resulting quartic equation admits only one positive solution for $\rho_0^2$. Thus, in contrast with the de Sitter case, only one physical branch is present. Moreover, setting $\dot{\rho}_{\text{AdS}}=0$ in the energy constraint yields a single positive turning point $\rho_*$. The allowed region is therefore $0<\rho_{\text{AdS}}\leq\rho_*$. Thus, the solution initially expands, reaches its maximum size at $\rho_*$, and then recollapses towards $\rho_{\text{AdS}}=0$. The latter corresponds to a Big-Crunch singularity, as evidenced by the Ricci scalar, which diverges as $\rho_{\text{AdS}}\rightarrow0$. Notice also that the characteristic frequency $\omega=\sqrt{-\Lambda/3}$ is independent of the number of flavors. Therefore, $N$ does not modify the characteristic time scale of the trigonometric solution, but affects its amplitude and the position of the turning point through the dependence of $b$ and $c$ on $a_N$. Hence, although $\rho_{\text{AdS}}^2(t)$ is formally periodic, its periodic continuation should not be interpreted as a regular oscillating cosmology, since the dynamical evolution finishes at the Big-Crunch singularity.

Finally, in the case of a vanishing cosmological constant, the energy constraint fixes the initial scale factor appearing in Eq.~\eqref{rho flat} to $\rho_0^2=b/2(1-c)$, which requires $c<1$, or equivalently $K\kappa a_N<2$. The evolution is restricted to a single allowed region, with one positive turning point at $\rho_*=\sqrt{2}\rho_0$. Hence, the solution initially expands, reaches a maximum size at $\rho_*$, and then recollapses towards a Big-Crunch singularity at $\rho_{\text{flat}}=0$. The number of flavors also affects the characteristic scales of this evolution. In particular, for fixed couplings and within the allowed parameter range, increasing $N$ increases both $\rho_0$ and $\rho_*$. However, notice that the condition $K\kappa a_N<2$ imposes an upper bound on the number of flavors allowed for fixed values of the couplings.\\

\subsection{Squashed Bianchi IX cosmology}

The metric in Eq. \eqref{Bianchi} can be generalized to a squashed spacetime by introducing a temporal squashing parameter, $\alpha(t)$, namely
\begin{equation} \label{Bianchi2}
    ds^2 =-dt^2 +\frac{\rho(t)^2}{4}\left[ \alpha(t)^2\left(d\gamma +\cos\theta d\phi\right)^2 +d\theta^2 +\sin^2 \theta d\phi^2\right] \ , 
\end{equation}
where the coordinate ranges are defined below Eq. \eqref{Bianchi}.

One can verify that within this ansatz the Skyrme equations remain identically satisfied, while the Einstein equations reduce to the following compatible system:
\begin{subequations} \label{squashing-eq}
\begin{gather} 
    2\alpha\rho^2\left(2\rho \dot{\alpha}+3\alpha\dot{\rho}\right)\dot{\rho}-2\alpha^2\rho^2\left(\Lambda \rho^2 +\alpha^2-4 \right)-a_{N}\kappa K\left[\left(2\rho^2 +\lambda \right)\alpha^2 +\rho^2 +2\lambda \right]=0 \, , \\
    2\alpha^2\rho^2 \left(2\rho \ddot{\rho}+\dot{\rho}^2 \right)-2\alpha^2\rho^2\left(\Lambda \rho^2 +3\alpha^2-4 \right)-a_{N}\kappa K\left[\left(2\rho^2+\lambda \right)\alpha^2-\rho^2-2\lambda \right]= 0 \, , \\
    \alpha\rho^3\left(\rho \ddot{\alpha}+3\dot{\rho}\dot{\alpha} \right)+\left(\alpha^2-1 \right)\left[a_{N}\kappa K \left(\lambda +\rho^2 \right)+4\alpha^2 \rho^2\right]=0 \, .
\end{gather}
\end{subequations}

The full solution to this system lies beyond the scope of this work.\footnote{For a full analysis of the dynamical squashing factor in the $SU(2)$ case, see Ref. \cite{Pavluchenko:2016gyd}.} In the linearized regime, to gain insight into the stability of the solution around the spherically symmetric case as a function of the flavor number $N$, we consider the expansions $\alpha(t)=1+\epsilon u(t)$ and $\rho(t)=\rho_{E}(t)+\epsilon v(t)$, where $\rho_{E}(t)$ denotes the undeformed solution corresponding to $\alpha(t)=1$. Substituting these expansions into Eqs. \eqref{squashing-eq}, the late-time behavior ($t\rightarrow \infty$) of both the scale factor and the squashing coefficient proves to be independent of the flavor number $N$, thus matching the results reported in Ref. \cite{Canfora:2017ivv}.

%%%%%%%%%%%%%%%%%%%%%%%%%%%
\section{Traversable Wormhole}    \label{Sec-V}
%%%%%%%%%%%%%%%%%%%%%%%%%%%

In this section, we show that the $SU(N)$ Einstein-NLSM admits exact solutions describing wormholes in an asymptotically AdS spacetime. The most remarkable feature of these solutions is that they possess timelike geodesics connecting both sides of the throat, meaning they are fully traversable. This solution was previously explored in Ref. \cite{Eloy1} for the $SU(2)$ case and subsequently extended to a non-trivial $SU(3)$ configuration in Ref. \cite{Eloy2}. Here, we generalize this construction to an arbitrary number of flavors $N$ and demonstrate that increasing the flavor number modifies the throat size, leading to desirable astrophysical features.

To construct these configurations, we draw inspiration from the cosmological solution presented in the previous section, which features a similar metric but includes a NUT parameter, $Q$,\footnote{This metric is obtained from Eq. \eqref{Bianchi} by performing a double Wick rotation. In this section, we use the coordinate $z$ instead of $\gamma$ to match the standard wormhole literature, although their coordinate ranges coincide, since $z$ represents the proper distance along the wormhole.} namely
\begin{equation} \label{metricWH}
    ds^2=dz^2+\frac{\rho^2(z)}{4}\left[-Q^2\left(dt+\cos\theta d\phi\right)^2+d\theta^2+\sin^2\theta d\phi^2\right]\,.
\end{equation}
In coordinates $x^{\mu}=\{t,z,\theta,\phi\}$, the matter field ansatz is given by
\begin{equation} \label{UWH}
    F_{1}(x^\mu)=t\,,\qquad F_{2}(x^{\mu})=\theta\,,\qquad F_{3}(x^{\mu})=\phi\,.
\end{equation}
It is straightforward to verify that within the ansatz defined by Eqs. \eqref{metricWH} and \eqref{UWH}, the Skyrme equations are identically satisfied for any value of $N$. On the other hand, the Einstein field equations yield three non-trivial components: the $zz$-component provides a first-order (energy) constraint, while the $\theta\theta$-component yields a second-order evolution equation for $\rho(z)$. The remaining $tt$-component is also a second-order equation for $\rho(z)$, which is consistent with the $\theta\theta$-equation only when the algebraic constraint
\begin{equation}
    (4Q^2-a_NK\kappa) \rho^2 -a_NK\kappa\lambda=0
\end{equation}
is satisfied. To fulfill this condition, we must set $\lambda=0$ and $Q^2=a_NK\kappa/4$. This restricts the system to the NLSM sector within a spacetime whose NUT parameter is fixed by the coupling constants of the theory.

Under these restrictions, the single non-trivial Einstein equation reduces to
\begin{equation} \label{EqrhoWH}
    (\rho')^2+\frac{\Lambda}{3}\rho^2+\frac{1}{4}(a_NK\kappa-8)=0\,,
\end{equation}
which yields the exact solution
\begin{equation} \label{rhoWH}
    \rho(z)=\rho_0\cosh\left(\sqrt{\frac{|\Lambda|}{3}}z\right)\,.
\end{equation}
Eq. \eqref{rhoWH} represents the shape function of the wormhole.

A few remarks regarding this solution are in order. In principle, Eq. \eqref{EqrhoWH} admits two distinct branches of solutions; furthermore, the cosmological constant can be either positive or negative, leading to different physical interpretations. However, a valid wormhole geometry imposes two requirements: First, spacetime regularity requires a negative cosmological constant, as the hyperbolic cosine in Eq. \eqref{rhoWH} prevents $\rho(z)$ from vanishing. This guarantees that all curvature invariants remain non-singular, as they depend on $\rho(z)$ and its inverse, both of which are strictly regular everywhere. Second, the negative solution branch of Eq. \eqref{EqrhoWH} is ruled out because the shape function must be strictly positive. Thus, Eq. \eqref{rhoWH} is the unique physically viable solution under the given constraints. The integration constant $\rho_0$ is given by
\begin{equation} \label{rho0_def}
    \rho_0= \sqrt{\frac{3\left(a_NK\kappa-8\right)}{4|\Lambda|}}\,,
\end{equation}
which is well-defined provided that $a_NK\kappa>8$.

This configuration represents a traversable Lorentzian AdS wormhole with a NUT parameter, bypassing well-known no-go theorems regarding wormhole existence \cite{Visser:1995cc,Friedman:1993ty}, where a negative cosmological constant serves as the required exotic source.

At this point, it is relevant to highlight that spacetimes with a NUT parameter are not unphysical curiosities. Recent developments have shown that such geometries represent a physically viable family of configurations \cite{Clement:2015cxa, Clement:2015aka}. Thus, the presence of a NUT parameter does not introduce pathologies, but rather places our multi-flavor solution within a well-behaved framework of traversable geometries.

The throat of the wormhole, defined as the region of minimum area connecting the two asymptotic regions of spacetime, corresponds to the global minimum of the shape function $\rho(z)$. The throat exists at a coordinate $z_0$ where $\rho'(z_0)=0$, provided that $\rho(z_0)>0$. Furthermore, to ensure the wormhole is traversable, the geometry must satisfy the flare-out condition, $\rho''(z_0)>0$. For our solution, the critical point occurs at $z_0=0$. Evaluating the second derivative, we find $\rho''(0)=\rho_0\frac{|\Lambda|}{3}>0$, confirming that the global minimum is located at the origin.

According to the metric in Eq. \eqref{metricWH}, the effective radius of the base space at any point $z$ is $R(z)= \rho(z)/2$. Consequently, the throat radius $R_0$ and the corresponding area $A$ are
\begin{equation}
    R_0=\frac{\rho_0}{2}=\sqrt{\frac{3\left(a_NK\kappa-8\right)}{16|\Lambda|}}\, , \qquad 
    A=4\pi R_0^2=\frac{3\pi\left(a_NK\kappa-8\right)}{4|\Lambda|}\,.
\end{equation}
From these results, we conclude that the wormhole throat expands when considering a greater number of flavors or, equivalently, when increasing the baryon number of the configuration.

Since enlarging the wormhole throat requires an increase in the matter content, a natural question arises regarding how the energy density scales with $N$ to sustain the throat; specifically, whether it grows indefinitely or stabilizes. To clarify this behavior, we compute the energy density of the configuration,
\begin{equation}
    \varepsilon=-T^t_t=\frac{2+a_NK\kappa}{\kappa\rho^2(z)}\,.
\end{equation}
Evaluating this expression at the throat ($z=0$) yields
\begin{equation}
    \varepsilon(z=0)=\frac{|\Lambda|}{3\kappa}\left(\frac{a_NK\kappa+2}{a_NK\kappa-8}\right)\,.
\end{equation}
Taking the large-$N$ limit gives
\begin{equation}
    \lim_{a_N\to\infty}\varepsilon(z=0)=\frac{|\Lambda|}{3\kappa}\,.
\end{equation}
This result shows that the energy density required to keep the throat open does not grow indefinitely with the number of flavors. Instead, it saturates at a finite value that depends exclusively on the cosmological constant and the gravitational coupling constant. Consequently, this configuration is efficient in the sense that the wormhole throat can be widened (by increasing $R_0$) without incurring a higher cost in terms of energy density at the throat itself.

To establish the traversability of this wormhole configuration, we examine the behavior of timelike geodesics traversing the throat region. Following the geodesic framework outlined in Ref. \cite{Eloy1} for the $SU(2)$ case, the effective Lagrangian for radial timelike curves parameterized by the proper time $s$ is given by $L_{\text{eff}}=g_{\mu\nu}\dot{q}^\mu\dot{q}^\nu$ with $q^\mu=(t(s),z(s),0,0)$. This dynamical setup possesses two fundamental constants of motion: the Lagrangian itself, satisfying $g_{\mu\nu}\dot{q}^\mu\dot{q}^\nu=-I^2$, and the conserved energy $E$ associated with the timelike Killing vector $\partial_t$. Incorporating these conserved quantities into the metric, the radial geodesic equation reduces to
\begin{equation}
    \left(\frac{dz}{ds}\right)^2-\frac{E^2}{Q^2\rho^2(z)}=-I^2\,.
\end{equation}
Remarkably, the qualitative and quantitative geodesic analysis demonstrated in Ref. \cite{Eloy1} remains fully valid for our general $SU(N)$ solution under the parametric rescaling $K\to a_NK$. The resulting trajectories describe bounded periodic motion confined within an attractive effective potential centered at $z=0$. For sufficiently small values of $I^2$, these test particles deeply penetrate and cross between both the positive-$z$ and negative-$z$ asymptotic AdS regions, demonstrating that the wormhole throat is traversable without encountering curvature singularities.

%%%%%%%%%%%%%%%%%%%%%%%%%%%%%%%%%%%%%
\section{Conclusions}   \label{Sec-VI}
%%%%%%%%%%%%%%%%%%%%%%%%%%%%%%%%%%%%%

In this article, we have investigated several aspects of exact, spherically symmetric solutions within the $SU(N)$ Einstein-Skyrme model. First, we have demonstrated that the theory supports hairy black hole solutions, whose thermodynamics are governed by a rich interplay between the number of flavors and the coupling constants of the theory. Second, traversable wormholes emerge as solutions to the NLSM sector; remarkably, extending to $N > 2$ flavors allows for a significant expansion of the wormhole throat, accompanied by a saturation of the energy density at the throat in the large-$N$ limit. Finally, isotropic Bianchi-IX cosmological solutions show that the flavor number modulates the global evolution by affecting turning points, transient expansion, and energy density. Specifically, while a higher flavor number slows down scale factor growth due to enhanced gravitational attraction, it transiently increases the expansion rate $H(t)$ via higher energy density. For $\Lambda>0$, this dynamics splits into a re-collapsing lower branch and an asymptotically de Sitter upper branch, whereas for $\Lambda\le0$, the evolution leads to a Big-Crunch singularity. Remarkably, these cosmological configurations possess a non-trivial topological charge, which identifies the matter content with baryonic states. Furthermore, the explicit dependence on the flavor number allows for an analytical exploration of the large-$N$ limit.

These findings suggest that the $SU(N)$ sector provides a versatile framework for exploring how internal symmetries and topological constraints fundamentally reshape the global structure of spacetime. Furthermore, our results open new avenues for investigating the stability and potential gravitational signatures of these configurations in more complex, non-stationary scenarios. Overall, these results underscore the robustness of the Einstein-Skyrme model in providing a consistent description of astrophysically relevant solutions.

%%%%%%%%%%%%%%%%%%%%%%%%%%%%%%%%%%%%%%%%%
\section*{Acknowledgments}

C. H. has been funded by FONDECYT postdoctoral Grant No. 3240632. A. V. has been funded by FONDECYT Iniciaci\'on No. 11261883. P. R. has been funded by FONDECYT postdoctoral Grant No. 3230633. P. M. and P. R. acknowledge the GIAMM Research Group at Universidad de Los Lagos for providing the institutional support that facilitated this work, and are grateful to the Centro Multidisciplinario de F\'isica of Universidad Mayor for its warm hospitality during the period in which part of this work was completed.

%%%%%%%%%%%%%%%%%%%%%%%%%%%%%%%%%%%%%%%%%

\end{document}